
%
\magnification=\magstep1
\input amstex
\vsize=22.5truecm
\hsize=16truecm
\hoffset=2.4truecm
\voffset=1.5truecm
\parskip=.2truecm
\font\ut=cmbx10
\font\ti=cmbx10 scaled\magstep1

\TagsOnRight
\newfam\rmfam
\def\rm{\fam\rmfam}
\font\tenrm = cmr10
\font\sevrm = cmr7 \skewchar\sevrm='177
\font\sevit = cmmi7 \skewchar\sevrm='177
\font\fivrm = cmr5 \skewchar\fivrm='177
\textfont\rmfam = \tenrm
\scriptfont\rmfam = \sevrm
\scriptscriptfont\rmfam = \fivrm

\def\al{\alpha}

\def\ga{\gamma}
\def\de{\delta}

\def\la{\lambda}

\def\Om{\Omega}

\def\vx{\vec{x}}
\def\vxp{\vec{x}\,'}
\def\vy{\vec{y}}

\def\vz{\vec{z}}
\def\vk{\vec{k}}

\def\vp{\vec{p}}

\def\vq{\vec{q}}


\def\od2{\overline{|\de_k|^2}}

\def\11r{|1,\overline{1};\Om\rangle}
\def\l11{\langle1,\overline{1};\Om|}

\def\vt{\!\triangle\!\!}



\def\kph{k_{\rm phys}}

\def\laph{\la_{\rm phys}}

\def\Hi{H_{\rm I}}
\def\Heq{H_{\rm EQ}}

\def\Mp{M_{\rm Pl}}


\def\rt{\rho_{\rm tot}}

\def\rmk{\rho_{\raise1pt\hbox{\sevrm mode}\,\,\vk}}
\def\rmp{\rho_{\lower1pt\hbox{\sevrm mode}\,\,\vp}}
\def\drmk{\de\rho_{\raise1pt\hbox{\sevrm mode}\,\,\vk}}
\def\drmq{\de\rho_{\lower1pt\hbox{\sevrm mode}\,\,\vq}}
\def\xx{\xi^{(2)}(\ell)}
\def\xxxLL{\xi^{(3)}(\ell,\ell';\cos\theta)}
\def\xxxT{\xi_{\,\vt}{}^{(3)}(\ell)}
\def\ddR{\langle\de_R^2\rangle}

\def\dddRR{\langle\de_R^3\rangle}
\def\dddRS{\langle\de_R^2\de_{\lower1pt\hbox{\sevit S}}\rangle}

\font \fivesans               = cmss10 at 5pt

\font \tenrm                  = cmr10
\font \sevensans              = cmss10 at 7pt
\font \tensans                = cmss10
\newfam\sansfam
\textfont\sansfam=\tensans\scriptfont\sansfam=\sevensans
\scriptscriptfont\sansfam=\fivesans
\def\sans{\fam\sansfam\tensans}



\def\bbbc{{\mathchoice {\setbox0=\hbox{$\displaystyle\hbox{\rm C}$}\hbox{\hbox
to0pt{\kern0.4\wd0\vrule height0.9\ht0\hss}\box0}}
{\setbox0=\hbox{$\textstyle\hbox{\rm C}$}\hbox{\hbox
to0pt{\kern0.4\wd0\vrule height0.9\ht0\hss}\box0}}
{\setbox0=\hbox{$\scriptstyle\hbox{\rm C}$}\hbox{\hbox
to0pt{\kern0.4\wd0\vrule height0.9\ht0\hss}\box0}}
{\setbox0=\hbox{$\scriptscriptstyle\hbox{\rm C}$}\hbox{\hbox
to0pt{\kern0.4\wd0\vrule height0.9\ht0\hss}\box0}}}}

\def\bbbq{{\mathchoice {\setbox0=\hbox{$\displaystyle\hbox{\rm Q}$}\hbox{\raise
0.15\ht0\hbox to0pt{\kern0.4\wd0\vrule height0.8\ht0\hss}\box0}}
{\setbox0=\hbox{$\textstyle\hbox{\rm Q}$}\hbox{\raise
0.15\ht0\hbox to0pt{\kern0.4\wd0\vrule height0.8\ht0\hss}\box0}}
{\setbox0=\hbox{$\scriptstyle\hbox{\rm Q}$}\hbox{\raise
0.15\ht0\hbox to0pt{\kern0.4\wd0\vrule height0.7\ht0\hss}\box0}}
{\setbox0=\hbox{$\scriptscriptstyle\hbox{\rm Q}$}\hbox{\raise
0.15\ht0\hbox to0pt{\kern0.4\wd0\vrule height0.7\ht0\hss}\box0}}}}

\def\bbbz{{\mathchoice {\hbox{$\sans\textstyle Z\kern-0.4em Z$}}
{\hbox{$\sans\textstyle Z\kern-0.4em Z$}}
{\hbox{$\sans\scriptstyle Z\kern-0.3em Z$}}
{\hbox{$\sans\scriptscriptstyle Z\kern-0.2em Z$}}}}
\def\qed{\ifmmode\sq\else{\unskip\nobreak\hfil
\penalty50\hskip1em\null\nobreak\hfil\sq
\parfillskip=0pt\finalhyphendemerits=0\endgraf}\fi}


\def\H{\Cal{H}}

\def\gati{\ga^{\lower2pt\hbox{$\ssize\tilde0$}}}

\rightline{ETH-TH/94--35}
\rightline{October 1994}
\vskip2truecm
\centerline{\ti Non--Gaussian Primordial Fluctuations }
\vskip0.2truecm
\centerline{\ti in Inflationary Cosmology}
\vskip2truecm
\centerline{Harald F. M\"uller and Christoph Schmid}
\vskip0.5truecm
\centerline{Institut f\"ur Theoretische Physik, ETH--H\"onggerberg,
            CH--8093 Z\"urich, Switzerland}

\vskip3.5truecm

\centerline{\ut Abstract}
\vskip0.5truecm
\noindent{\sl
 We analyze the non--Gaussian primordial fluctuations which are inescapably
 contributed by scalar fields $\Phi$ with vanishing expectation values,
 $\langle\Phi\rangle=0$, present during inflation in addition to the
 inflaton field. For simplicity we take $\Phi$ to be non--interacting and
 minimally coupled to gravity. $\Phi$ is a Gaussian variable, but the energy
 density fluctuations contributed by such a field are $\chi^2$--distributed.
 We compute the three--point function $\xxxT$ for the configuration of an
 equilateral triangle (with side length $\ell$) and the skewness $\dddRR$,
 {\it i.e.} the third moment of the one--point probability distribution of
 the spatially smeared energy density contrast $\de_R$, where $R$ is the
 smearing scale. The relative magnitudes of the non--Gaussian effects,
 $[\xi^{(N)}]^{1/N}/[\xi^{(2)}]^{1/2}$, do not grow in time. They are given
 by numerical constants of order unity, independent of the scale $\ell$.
 The "bi--skewness" $\dddRS$ is positive. For smearing lengths $R\ll S$
 this shows that in our model (in contrast to Gaussian models) voids are
 more quiet than high--density regions.
}


\vfill\eject

\noindent
{\ti 1. Introduction and Conclusions}
\vskip0.5truecm

Gaussian primordial fluctuations are predicted by standard inflationary
models [1]. The data on the anisotropies of the cosmic microwave
background taken by the COBE satellite [2] are not sufficient to give a
significant test for this prediction of a Gaussian distribution [3].
Large non--Gaussian effects are observed in the large--scale structure of
the galaxy distribution, {\it e.g.} in the one--point probability
distribution of the matter density $\rho$, smeared over a window scale $R$,
where $R$ is in the range 5 to 20 $h^{-1}$ {\sl Mpc} [4]. One inevitable
source for this non--Gaussianity is the non--linear character of the
gravitational evolution, irrespective if the initial fluctuations are
Gaussian or not.

To test the hypothesis of Gaussian primordial fluctuations, one needs one
or several specific non--Gaussian models. One class of models which give
rise to non--Gaussian primordial fluctuations involves cosmological defects,
like cosmic strings or textures [5].

In this paper we discuss a different origin of non--Gaussianity in the
primordial fluctations, and we analyze a minimal non--Gaussian model, which
has been introduced in ref. [6]. Non--Gaussian primordial fluctuations are
inescapably contributed by scalar fields $\Phi$ with vanishing expectation
value, $\langle\Phi\rangle=0$, which are present during inflation in
addition to the inflaton field. For simplicity we consider $\Phi$ to be a
non--interacting scalar field, minimally coupled to gravity. $\Phi$ is a
Gaussian variable. Since the energy density $\rho$ of this non--inflaton
field is bilinear in the fluctuating field $\Phi$, $\rho$ is
$\chi^2$--distributed. (Recall that the $\chi^2$--distribution is defined as
the probability distribution of the mean square error of a Gaussian
variable). In contrast the energy density fluctuations of the inflaton arise
from the interference terms between the background field and the fluctuating
part of the inflaton field. Therefore the energy density fluctuations are
linear in the inflaton field fluctuations, and the inflaton energy density
fluctuations are Gaussian.

The order of magnitude of the fluctuations at the second horizon crossing
contributed by non--inflaton massless scalar fields is
$\de\rho/\rho\sim\Hi^2/\Mp^2$, where $\Hi$ is the Hubble constant during
inflation. For massive, unstable scalar fields with masses $m_\Phi\ll\Hi$,
$\de\rho/\rho$ acquires an enhancement factor $\sqrt{m_\Phi/\Gamma_\Phi}$,
where $\Gamma_\Phi$ is the decay rate [7]. In this short note we consider
only the massless case, in refs. [6,7] we have analyzed both $m_\Phi=0$
and $m_\Phi\neq0$. In previous analyses of non--Gaussian fluctuations in
inflationary models with two scalar fields [8] the physical origin and the
resulting character of the non--Gaussianity are very different from the ones
presented here.

Observable measures of non--Gaussianity are the expectation values of
products of $N$ energy density contrasts $\de(\vx)$ for $N>2$. In this
paper we restrict ourselves to $N=3$. We first (in sect. 2) compute the
three--point function
$\xi^{(3)}(\vx_1,\vx_2,\vx_3)=\langle\de(\vx_1)\de(\vx_2)\de(\vx_3)\rangle$
for the configuration of an equilateral triangle with sides of length
$\ell$, {\it i.e.} $\xxxT$. As a second observable we compute (in sect. 3)
the skewness $\dddRR$, {\it i.e.} the third moment of the one--point
probability distribution of the spatially smeared energy density contrast
$\de_R:=\int d^3x\,\,\de(\vx)W_R(\vx)$ with a window function $W_R(\vx)$ of
smearing scale $R$. As a third measure of non--Gaussianity we consider (in
sect. 4) the "bi--skewness" $\dddRS$ in the case $R\ll S$.

The non--Gaussianity of the primordial fluctuations in our model has a
simple structure. For a $\chi^2$--distributed variable, such as the energy
density contrast $\de$, the expectation value of a product of $N$ $\de$'s is
of the order of the $(N/2)^{th}$ power of the expectation value of two
$\de$'s, {\it e.g.} for the $N$--point function we find a $\xi^{(N)}$ of the
order of $[\xi^{(2)}]^{N/2}$. This is so because any $\xi^{(N)}$ must be
built out of products of $N/2$ two--point functions of $\Phi$.

The relative magnitude of the non--Gaussian effects in $\xi^{(N)}$ is given
by \linebreak
$[\xi^{(N)}]^{1/N}/[\xi^{(2)}]^{1/2}$. During the cosmological time
evolution the perturbations $\xi^{(2)}$ grow for fixed comoving $\ell$. But
in our class of models (with its linear dynamics) the ratios
$[\xi^{(N)}]^{1/N}/[\xi^{(2)}]^{1/2}$ are independent of time. They are
given by numerical constants of order unity.

The cosmological model considered is an inflationary de Sitter era, followed
by a radiation dominated era. The quantum state of the field $\Phi$ is the
Bunch--Davies state initially, during inflation. The scalar field $\Phi$
(not the inflaton) evolves in this background curved space--time, {\it i.e.}
the back reaction of $\Phi$ on the geometry is neglected. Therefore gauge
ambiguities are eliminated [9]. Since we do not treat the decay of the
scalar field (into radiation or other forms of matter), we make our
predictions for the end of the radiation dominated era.

At the end of the radiation era (the time of matter and radiation equality)
the model contains two characteristic scales, the Hubble parameters $\Hi$
and $\Heq$. For cosmologically relevant length scales the quantities
$\xxxT$ and $\dddRR$ each contain two domains according to whether $\ell$ or
$R$ are smaller or bigger than $\Heq$. On sub--horizon scales,
$\Heq\ell\ll1$, the three--point function increases logarithmically towards
the smaller scales, $\xxxT\sim|\log\H\ell|^3$. On super--horizon scales,
$\Heq\ell\gg1$, we obtain $\xxxT\sim(\H\ell)^{-6}$. We find the scaling laws
$\xi^{(3)}/[\xi^{(2)}]^{3/2}=5/\sqrt{6}$ for sub--horizon and $=\sqrt{8/27}$
for super--horizon scales. We obtain $\dddRR/\ddR^{3/2}=10/(6\pi e^3)^{1/4}$
for the ratio of skewness and variance of $\de_R$ with smearing scales
$\Heq R\ll1$. The "bi--skewness" $\dddRS$ is positive for all scales $R$ and
$S$. For $R\ll S$ this shows that in our model the small--scale variance
$\ddR$ observed in a sample within a void of size $S\gg R$ is expected to be
smaller than the small--scale variance observed in a high--density region of
size $S$.

The non--linear gravitational evolution of Gaussian primordial fluctuations
produces another simple relation between expectation values of $N$ $\de$'s
in lowest non--vanishing order perturbation theory [10]. In this approach
the $N$--point functions $\xi^{(N)}$ are of the order $[\xi^{(2)}]^{N-1}$,
completely different from the relation valid for a $\chi^2$--distributed
quantity.

According to refs. [11,12] the observed $N$--point correlation functions of
galaxy counts for $N=3,4$ in the range 1 to 50 $h^{-1}$ {\sl Mpc} are
consistent with non--linear gravitational evolution from Gaussian primordial
fluctuations. In order to see how much room the observations leave for
non--Gaussian primordial fluctuations, it would be necessary to apply the
non--linear gravitational evolution to these fluctuations.

\vskip1truecm

\noindent
{\ti 2. The Three--Point Correlation Function}
\vskip0.5truecm

The background geometry is a Friedmann--Robertson--Walker space--time with
spatially flat sections, $\rt=\rho_{\rm crit}$. With conformal time $\eta$
we have $ds^2=a(\eta)^2(d\eta^2-dx_idx^i)$. The Hubble constant during
inflation is denoted by $\Hi$. The transition between inflation and the
radiation era can be approximated as instantaneous, because the physical
transition time is much shorter than the characteristic time for the
evolution of the cosmologically relevant modes, which have
$\laph>e^{70}\Hi^{-1}$. We fix the coordinates in such a way that at the
time $\eta_1$ of the transition ({\it i.e.} at the end of inflation) we have
$\eta_1=\Hi{}^{-1}$ and $a(\eta_1)=1$. This fixes the scale factor as
$$
\alignedat2
a(\eta)\,\,&=\,\,{1\over(2-\Hi\eta)}&&\quad(\eta\le\Hi^{-1}) \\
a(\eta)\,\,&=\,\,\Hi\eta            &&\quad(\eta\ge\Hi^{-1}).
\endalignedat
\tag2.1
$$
The Hubble parameter $\H(\eta)$ during the radiation era reads
$\H=a^{-2}\Hi$.

We consider a non--interacting, massless, neutral scalar field $\Phi$ with
minimal coupling to gravity. Its energy density $\rho$ is
$$
\rho\,\,=\,\,{1\over2a^2}\,\bigg[(\partial_\eta\Phi)^2 +
                                               (\partial_{\vx}\Phi)^2\bigg].
\tag2.2
$$
The energy density contrast contributed by the scalar field considered is
$$
\de(x)\,\,=\,\,{1\over\rt}(\rho(x)-\langle\rho\rangle),
\tag2.3
$$
where $\rt=3\H^2\Mp^2/8\pi$.

The equal time three--point function of the energy density contrast is
$$
\xi^{(3)}(\vx_1,\vx_2,\vx_3)\,\,=\,\,
                   \langle\Om|\,\de(\vx_1)\de(\vx_2)\de(\vx_3)\,|\Om\rangle.
\tag2.4
$$
The quantum state $|\Om\rangle$ is initially (during inflation) the
Bunch--Davies state: Every observationally relevant mode, which has today
$(\laph)_0<H_0^{-1}$, had at early times in inflation $R/\kph{}^2\to0$ and
is taken to be initially in the Minkowski vacuum state. Since the quantum
state $|\Om\rangle$ is translation invariant, $\xi^{(3)}$ depends only on
the mutual physical separations of the three points $\vx_i$.

We work with a normal ordered energy density operator N$[\rho]$, and we use
the fact that any normal ordering N gives the same result for the
correlation functions. This is so because the difference of two normal
orderings is a c--number, which drops out in the density contrast, eq.
(2.3), and a fortiori in the correlation functions.

Since the results are independent of the normal ordering, we can make the
most convenient choice, the one adapted to the quantum state
$|\Omega\rangle$. For the mode expansion of the field operator $\Phi$,
$$
\Phi(\eta,\vx)\,\,=\,\,\int{d^3k\over(2\pi)^3}\,\,
   \big[\,a_{\vk}\,\varphi_{\vk}(\eta,\vx)\,+\,
          a_{\vk}{}^\dag\,\varphi_{\vk}(\eta,\vx)^*\,\big],
\tag2.5
$$
we choose the expansion in terms of those evolving basis modes
$\varphi_{\vk}(\eta,\vx)$ which at very early times, when
$\laph\ll\Hi^{-1}$, behaved as $\sim\exp(i\vk\vx-ik\eta)$. It follows that
$$
a_{\vk}|\Omega\rangle\,\,=\,\,0,\quad\forall\vk.
\tag2.6
$$
One could write more explicitely
$\varphi_{\vk}(\eta,\vx)=\varphi_{\vk}{}^{\rm (in)}(\eta,\vx)$,
$a_{\vk}=a_{\vk}{}^{\rm (in)}$, and $|\Om\rangle=|BD,\,\rm in\rangle$ (where
$BD$ stands for Bunch--Davies). For more details about the physics and the
functional form of the evolved modes we refer to refs. [6,7].

In [6] it was shown that the Wightman function $W(x,x')$ of the state
$|\Om\rangle$,
$$
W(x,x')\,:=\,\,\langle\Om|\Phi(x)\Phi(x')|\Om\rangle\,\,=\,\,
\int{d^3k\over(2\pi)^3}\,\,\varphi_{\vk}(x)\varphi_{\vk}(x')^*,
\tag2.7
$$
is extremely useful, because the equal time two--point function of the
energy density contrast
$\xi^{(2)}(\vx,\vxp)=\langle\Om|\de(\vx)\de(\vxp)|\Om\rangle$ in the state
$|\Om\rangle$ can be expressed in terms of derivatives of $W(x,x')$ as
$$
\xi^{(2)}(\vx_1,\vx_2)\,\,=\,\, {2\over\big(2a^2\rt\big)^2}\,
     \sum_{\al_1,\al_2=0}^3\big[W(x_1,x_2)_{,\al_1\al_2}\big]^2
                                            \bigg|_{\dsize\eta_1=\eta_2}.
\tag2.8
$$
The derivatives with index $i$ act exclusively on the coordinates of the
point $x_i$. For the three--point function $\xi^{(3)}$ the analogous
expression reads
$$
\aligned
&\xi^{(3)}(\vx_1,\vx_2,\vx_3)\,\,=\\
 &{8\over\big(2a^2\rt\big)^3}\,
     \sum_{\al_1,\al_2,\al_3=0}^3W(x_1,x_2)_{,\al_1\al_2}\cdot
          W(x_2,x_3)_{,\al_2\al_3}\cdot W(x_3,x_1)_{,\al_3\al_1}
                                    \bigg|_{\dsize\eta_1=\eta_2=\eta_3}.
\endaligned
\tag2.9
$$

Cosmologically relevant separations of the three points, today 1 to 3000
$h^{-1}$ {\sl Mpc}, correspond to physical point separations at the end
of inflation much larger than the Hubble radius $\Hi^{-1}$ at that time.
Therefore we can use an approximate Wightman function for the state
$|\Om\rangle$ during the radiation era [7],
$$
\aligned
W(x,x')\,\,=\,\,
{\Hi^2\over(2\pi)^2}{1\over24\eta\eta'\vt r}\,
\bigg[&\,\Delta_{-+}^3\,\log |\Hi\Delta_{-+}|\,\,+\,\,
         \Delta_{+-}^3\,\log |\Hi\Delta_{+-}| \\
   &-\,\,\Delta_{--}^3\,\log |\Hi\Delta_{--}|\,\,-\,\,
         \Delta_{++}^3\,\log |\Hi\Delta_{++}|\,\,\bigg].
\endaligned
\tag2.10
$$
The comoving distance of the two points is denoted by $\vt r:=|\vx-\vxp|$,
and we have introduced the notation
$\Delta_{\pm\pm}:=\,\,\vt r\pm\eta\pm\eta'$.

The three points $\vx_i$ in eq. (2.4) define the corners of a triangle.
We consider the special case where the three points form an equilateral
triangle, and we denote the corresponding three--point correlation function
by $\xxxT$, where $\ell$ is the physical distance between any two of the
points. The model contains two intrinsic length scales, namely $\Hi^{-1}$
and $\H^{-1}$. For cosmologically relevant point separations,
$\ell\gg a\Hi^{-1}$, the three--point function contains two domains,
according to whether $\ell$ is bigger or smaller than $\H^{-1}$. For
sub--horizon length scales, $\H\ell\ll1$, we find
$$
\xxxT\,\,\simeq\,\,
 {5\over36}\bigg({2\over3\pi}\bigg)^3\,\bigg({\Hi\over\Mp}\bigg)^6\,
                                           \bigg|\log{\H\ell\over2}\bigg|^3.
\tag2.11
$$
On super--horizon scales, $\H\ell\gg1$, we obtain
$$
\xxxT\,\,\simeq\,\,
 \bigg({2\over3\pi}\bigg)^3\,\bigg({\Hi\over\Mp}\bigg)^6\,
                                                        {1\over(\H\ell)^6}.
\tag2.12
$$

Comparing eq. (2.8) for $\xi^{(2)}$ with eq. (2.9) for $\xi^{(3)}$ we expect
that there should be a scaling law $\xxxT=\Cal{O}[\xx]^{3/2}$. Using the
results of ref. [7] for $\xx$ we find
$$
{\xxxT\over[\xx]^{3/2}}\,\,=\,\,
\left\{\aligned
         {5\over\sqrt{6}} \quad&{\rm for}\,\,\ell\ll\H^{-1} \\
         \sqrt{{8\over27}}\quad&{\rm for}\,\,\ell\gg\H^{-1}.
       \endaligned
\right.
\tag2.13
$$
On sub--horizon $(\H\ell\ll1)$ and super--horizon scales $(\H\ell\gg1)$ we
have scaling laws. The constant coefficients are different in the two
regimes, such that in the vicinity of $\ell=\H^{-1}$ the ratio
$\xxxT/[\xx]^{3/2}$ is slightly $\ell$--dependent, see figure.

\vskip1truecm

\noindent
{\ti 3. The Skewness}
\vskip0.5truecm

We introduce the spatially smeared energy density contrast
$$
\de_R\,:=\,\,\int d^3x\,\,\de(\vx)\,W_R(\vx).
\tag3.1
$$
For convenience, we take a Gaussian window function $W_R$ of scale $R$,
$$
W_R(\vx)\,\,=\,\,{1\over(\sqrt{2\pi}\,R)^3}\,e^{\tsize-{x^2\over2R^2}}.
\tag3.2
$$
We compute the second and third moments, $\ddR$ and $\dddRR$, of the
one--point probability distribution for $\de_R$. Note that one--point
probability distributions inescapably need a smearing scale $R$, both in the
analysis of observed galaxies and in quantum field theoretic computations.
The reason is that the observed mass density in galaxy counts is the sum of
Dirac delta functions (galaxies are treated as points), which cannot be
taken to the $N^{th}$ power without smearing beforehand. In quantum field
theory, on the other hand, the universal short--distance behaviour,
{\it e.g.} $\xx\sim\ell^{-8}$, necessitates smearing before squaring. The
cosmologically relevant term has a short--distance behaviour
$\sim(\log\H\ell)^2$, which also necessitates smearing.

The variance $\ddR=\int d^3x\int d^3y\,W_R(\vx)W_R(\vy)\,\xi^{(2)}(\vx,\vy)$
is given by a double integral. Since the equal time two--point function
$\xi^{(2)}$ can depend only on the physical separation $\ell$ of the two
points, one integration is trivial. We obtain
$$
\ddR\,\,=\,\,{1\over2\sqrt{\pi}R^3}\,\int_0^\infty {d\ell\over\ell}\,
                              \ell^3\,\,e^{\tsize-{\ell^2\over4R^2}}\,\,\xx.
\tag3.3
$$
The skewness $\dddRR=\int d^3x\int d^3y\int d^3z\,W_R(\vx)W_R(\vy)W_R(\vz)
                                                   \,\xi^{(3)}(\vx,\vy,\vz)$
is given as a triple integral. The equal time three--point function
$\xi^{(3)}$, which depends only on the mutual physical separations of $\vx$,
$\vy$, and $\vz$, can be characterized by any three quantities defining the
triangle given by the points. We choose two sides of length $\ell$ and
$\ell'$ and the angle $\theta$ in between, $\xxxLL$. One integration is
again trivial, and we arrive at
$$
\dddRR\,\,=\,\,{1\over\sqrt{27}\pi R^6}\,
 \int_0^\infty {d\ell\over\ell}\int_0^\infty {d\ell'\over\ell'}
   \int_{-1}^1 d(\cos\theta)\,\,\ell^3{\ell'}^3\,\,
         e^{\tsize-{\ell^2+{\ell'}^2-\ell\ell'\cos\theta\over3R^2}}\,\,\xxxLL.
\tag3.4
$$

Let us first concentrate on the case of sub--horizon smearing scales,
$\H R\ll1$. For sub--horizon separations, $\ell,\ell'\ll\H^{-1}$, the
three--point correlation function varies logarithmically,
$\xxxLL\simeq{5\over36}\Hi^6\Mp^{-6}(2/3\pi)^3
                                       |\log\H\ell\log\H\ell'\log\H\ell''|$,
where \linebreak
$\ell''=|\vec{\ell}-\vec{\ell}\,'|$. The integral (3.4) therefore is
strongly peaked at values $\ell=\ell'=3R/\sqrt{2-\cos\theta}$, and we obtain
$$
\dddRR\,\,\simeq\,\,{5\sqrt{3}\over\pi e^3}\,\bigg({2\over3\pi}\bigg)^3\,
                       \bigg({\Hi\over\Mp}\bigg)^6\,\,\big|\log\H R\big|^3
\quad (\H R\ll1).
\tag3.5
$$
Comparing the variance [7] and skewness we obtain the numerical constant
in the general scaling law discussed in the introduction,
$$
{\dddRR\over[\ddR]^{3/2}}\,\,=\,\,10\,
                      \root{\raise2pt\hbox{\sevrm 4}}\of{{1\over6\pi e^3}}
\qquad (\H R\ll1).
\tag3.6
$$

For super--horizon smearing scales, $\H R\gg1$, the exponential in eq.
(3.4) can be dropped, and the skewness $\dddRR$ is approximately
$$
\dddRR\,\,\simeq\,\,{8\sqrt{3}\over81\pi}\,C_3\,\bigg({\Hi\over\Mp}\bigg)^6\,
 {1\over(\H R)^6}\quad(\H R\gg1),
\tag3.7a
$$
where the numerical constant $C_3$ is defined by the integral of
$\xi^{(3)}$,
$$
\int d^3\ell\int d^3\ell'\,\,\xi^{(3)}(\ell,\ell',\ell'')\,\,=:\,
   \bigg({8\pi\over3}\bigg)^3\,\bigg({\Hi\over\Mp}\bigg)^6\,{C_3\over\H^6}.
\tag3.7b
$$
For $\H R\gg1$ the general scaling law is violated. The reason for this is
simple: If the physical separations of the points involved in $\xi^{(2,3)}$
are larger than $\H^{-1}$, both correlation functions show a steep
fall--off, $\xx\sim(\H\ell)^{-4}$ and $\xxxLL\sim(\H^3\ell\ell'\ell'')^{-2}$.
The integrals (3.3) and (3.4) are therefore strongly peaked at values for
$\ell$, $\ell'$, and $\ell''$ of the order of $\H^{-1}$, independent of the
scale $R\gg\H^{-1}$. But the $R$--dependence from the prefactors in eqs.
(3.3) and (3.4) gives a scaling behaviour such that another ratio,
$\dddRR/[\ddR]^2$, is independent of $R$. This reflects the fact that $\ddR$
involves three non--trivial integrations, while for $\dddRR$ six are needed.

The fact that the skewness is positive in our model (with its linear
dynamics) is totally independent of the kinematical effect, $\rho\ge0$,
$\de\ge-1$, which forces large fluctuations (needing non--linear dynamics)
to be associated with positive skewness.

\vskip1truecm

\noindent
{\ti 4. The Bi--Skewness}
\vskip0.5truecm

In the analysis of expectation values of three observables in position
space, we now consider, after $\xxxT$ and $\dddRR$, the "bi--skewness"
$\dddRS$, where the two smearing scales $R$ and $S$ (for density contrasts
at the same point) are different. This measure of non--Gaussian effects
appears to be both robust (observationally) and discriminating (with
respect to models). For scales $R\ll S$ the bi--skewness $\dddRS$ provides
the answer to the following physical question: Is the small--scale variance
$\ddR$ observed in a sample within a void of size $S$ ($S\gg R$) expected to
be smaller, equal, or larger than the small--scale variance observed within
an extended high density region of size $S$ ? For the cosmic microwave
background anisotropies the physical question is: Are the small--angle ($R$)
anisotropies observed within a moderately large patch ($S$) over the South
Pole expected to be the same as those observed within a moderately large
patch ($S$) over the North Pole, in view of the fact that the average
temperatures over the two different samples are different ? For Gaussian
fluctuations the small--scale variance is expected to be the same for
samples within large voids and for samples within large high--density
regions. For our model of non--Gaussianity (with its linear dynamics), the
voids are more quiet than the high--density regions (with respect to the
small--scale variance). For $R\ll S$ we find $\dddRS>0$. We obtain the
integral expression in the limit $R\ll S$
$$
\aligned
\dddRS\,\,=\,\,{1\over\sqrt{8}\pi R^3S^3}\,
 \int_0^\infty {d\ell\over\ell}\int_0^\infty {d\ell'\over\ell'}
   \int_{-1}^1 d(cos\theta)\,\,\ell^3{\ell'}^3\,\,e^{\tsize-{\ell^2\over4R^2}}
                    \,\,e&^{\tsize-{{\ell'}^2-\ell\ell'\cos\theta\over2S^2}}\\
                   \times&\xxxLL.
\endaligned
\tag4.1
$$
We insert the Wightman function (2.10), and for sub--horizon smearing
scales, \linebreak
$R\ll S\ll\H^{-1}$, we obtain
$$
\dddRS\,\,\simeq\,\,{15\over2\pi e^3}\,\bigg({2\over3\pi}\bigg)^3\,
    \bigg({\Hi\over\Mp}\bigg)^6\,\,\big|\log\H R\big|\big(\log\H S\big)^2.
\tag4.2
$$
The physics reason why the small--scale ($R$) variance is smaller in samples
(of size $S\gg R$) from voids than in samples from overdense regions is
easily seen, when we consider only two modes for simplicity, one mode with
$k^{-1}\approx R$, the other mode with $k^{-1}$ of the order of the
separation of the two samples. Since $\de\rho$ contains interference effects
between the two modes, the void is less noisy than the overdense region.

In the opposite limit, when the two scales obey $R\gg S$, the bi--skewness
$\dddRS$ is an uninteresting quantity, since it contains no new information
in addition to the skewness $\dddRR$.

\vskip1truecm

\noindent{\ti References}
\vskip0.5truecm

\item{[1]}  For a review see {\it e.g.}
            E.W. Kolb and M.S. Turner, {\sl The Early Universe},
            Addison--Wesley, 1990;
            V.F. Mukhanov, H.A. Feldman, and R.H. Brandenberger,
            Phys. Rep. {\bf215}(1992), 203.
\item{[2]}  G. Hinshaw, {\it et al.}, COBE--preprint (November 1993),
            astro-ph/9311030.
\item{[3]}  M. Srednicki, Ap. J. Lett. {\bf416}(1993), L1.
\item{[4]}  L. Kofman, E. Bertschinger, J.M. Gelb, A. Nusser, and A. Dekel,
            Ap. J. {\bf420}(1994), 44.
\item{[5]}  D. Coulson, P. Ferreira, P. Graham, and N. Turok, Nature
            {\bf 368}(1994), 27.
\item{[6]}  H.F. M\"uller and C. Schmid, ETH--preprint ETH--TH/93--15
            (January 1994), \linebreak
            gr-qc/9401020, {\it submitted} to Phys. Rev. D.
\item{[7]}  H.F. M\"uller and C. Schmid, ETH--preprint ETH--TH/94--25
            ({\it in preparation});\linebreak
            H.F. M\"uller, ETH--dissertation No. 10727 (1994), 125 pp.
\item{[8]}  S. Mollerach, S. Matarrese, A. Ortolan, and F. Luchin,
            Phys. Rev. D {\bf44}(1991), 1670;
            T.J. Allen, B. Grinstein, and M.B. Wise,
            Phys. Lett. {\bf 197B}(1987), 66.
\item{[9]}  J.M. Bardeen, P.J. Steinhardt, and M.S. Turner,
            Phys. Rev. D {\bf28}(1982), 679.
\item{[10]} J.N. Fry, Ap. J. {\bf 279}(1984), 499.
\item{[11]} D.J. Baumgart and J.N. Fry, Ap. J. {\bf 375}(1991), 25.
\item{[12]} F.R. Bouchet, M.A. Strauss, M. Davis, K.B. Fisher, A. Yahil, and
            J.P. Huchra, \linebreak Ap. J. {\bf417}(1993), 36.

\end